\documentclass[epj]{svjour}
\usepackage{amssymb,color}
\usepackage{graphics}
\usepackage[utf8]{inputenc}
\usepackage{graphicx,colordvi}
\usepackage{amsmath,amssymb}
\usepackage{color}   
\begin{document}
\title{Shape isomers in Pt, Hg and Pb isotopes with  N $\le$ 126}

\author{K. Pomorski, B. Nerlo-Pomorska, A. Dobrowolski, J. Bartel$^*$,
C. M. Petrache$^{**}$}
\institute{UMCS, Lublin, Poland,{$^*$} IPHC, CNRS-IN2P3, Strasbourg, France,
    {$^{**}$} CSNSM, CNRS-IN2P3, Orsay, France}

\date{Received: October 8, 2019 / Revised version: \today}

\abstract{Deformation-energy surfaces of 54 even-even isotopes of Pt, Hg and Pb
nuclei  with neutron numbers up to 126 are investigated within a
macroscopic-microscopic model based on the Lublin-Strasbourg-Drop macroscopic
energy and shell plus  pairing-energy corrections obtained from a Yukawa-folded
mean-field potential  at the desired deformation. A new, rapidly converging
Fourier shape  parametrization is used to describe nuclear shapes. The stability
of shape  isomeric states with respect to non-axial and higher-order
deformations is  investigated.}
\maketitle 
 

\section{Introduction} 
\label{intro} 

Recent experimental investigations studying the structure of neutron deficient
Lead and Mercury isotopes evidenced a shape coexistence phenomenon (see e.g.\
Refs.~\cite{And00,HWo11}) thus giving a strong impetus to theoreticians to
explore this region of nuclei in more detail. Important advances in the study of
neutron-deficient $Z\approx82$ nuclei have been realized using tagging
techniques in the Accelerator Laboratory of the University of Jyv\"askyl\"a,
Finland~\cite{Julin2016},  through Coulomb-excitation experiments undertaken at
the REX-ISOLDE facility in CERN~\cite{WLG16}, or isomeric and $\beta$-decays of
nuclei populated by relativistic energy-fragmentation experiments
\cite{podo2012} performed at GSI within the RISING campaign. Several interesting
theoretical studies on that subject, based either on a  self-consistent
approach  (see e.g.\ Refs.\ \cite{BHR03,YBH13,NRR13,NVR02}) or on
macroscopic-microscopic models (see e.g.\ \cite{MSB09}), have been published in
the last two decades.

The aim of the current paper is to present a systematic study of the potential
energy surfaces of even-even Pt, Hg, and Pb isotopes with neutron number $N\leq
126$ using a very efficient shape parametrization that we recently developed 
\cite{PNB15,SPN17}. We have chosen to carry out our study using the
macroscopic-microscopic (mic-mac) approach with the Lublin-Strasbourg Drop 
(LSD) \cite{PDu03}, able to reproduce accurately both nuclear masses and fission
barriers, and a Yukawa-folded single-particle potential \cite{DNi76,DPB16} to
account for shell and pairing-energy corrections.   A similar kind of
investigation of nuclear potential-energy surfaces for a wide  range of nuclei
has been carried out in Ref.~\cite{MSB12} using the 
($\varepsilon_2,\,\gamma,\,\varepsilon_4$) deformation space. Our study is, 
however, different in several aspects:  another expression is used for the
macroscopic energy, the evaluation of  pairing and shell-correction energy is
different, and, most importantly, a  better more general nuclear shape
parametrization is used in our present  approach. In addition, the range of
nuclear quadrupole deformations in Ref. \cite{MSB12} is smaller than the one in
our present study, with the consequence that the super-deformed minima found by
us are not present in the energy landscapes presented in Ref.~\cite{MSB12}.

The present investigation is also similar to those described in our  previous
study on shape isomers in the same mass region \cite{NPB17}, with,  however,
several differences:
                                                                     \\[ 0.5ex]
$\bullet$ 
                                                                    \\[ -1.8ex]
${}\hspace{0.2cm}$
\parbox{8.2cm}{the present calculations are carried out for a larger set of 
 isotopes ($^{170-204}$Pt, $^{172-206}$Hg, and $^{174-208}$Pb),}
                                                                    \\[ 0.5ex]
$\bullet$ 
                                                                    \\[ -1.8ex]
${}\hspace{0.2cm}$
\parbox{8.2cm}{the potential energy surfaces (PES) are studied here in two 
4-dimensional 
 deformation spaces: ($q_2,\,q_3,\,q_4,\,q_6$) and ($q_2,\,q_3,\,q_4,\,\eta$), 
 where $\eta$ describes a non-axial degree of freedom, and the $q_i$ 
 parameters correspond roughly to multipole deformations,}
                                  \\[ 0.5ex]
$\bullet$ 
                                                                    \\[ -1.7ex]
${}\hspace{0.2cm}$
\parbox{8.2cm}{a denser grid of deformation parameters than that of 
Ref.\ \cite{NPB17} is used,}
                                  \\[ 0.5ex]
$\bullet$ 
                                                                    \\[ -1.7ex]
${}\hspace{0.2cm}$
\parbox{8.2cm}{an average pairing energy taking into account the effect of an 
 approximate particle-number projection is included.}


\section{Theoretical model and details of the calculations}
\label{Sect.02}

In the macroscopic-microscopic method \cite{MSw66} the total energy of a 
nucleus at a given deformation can be calculated as the sum of the macroscopic 
(liquid-drop type) energy and quantum correction terms for protons and neutrons 
generated by shell and pairing effects:
\begin{equation}
 E_{\rm tot} = E^{}_{\rm LSD}+ {\rm E_{shell}} + {\rm E_{\rm pair}} \; .
\label{eq01}\end{equation}
The LSD model \cite{PDu03} is used to evaluate the macroscopic part of the
energy. Shell corrections are obtained, as usual, by subtracting the average
energy $\widetilde E$ from the sum of the single-particle (s.p.) energies of
occupied orbitals
\begin{equation}
 {\rm E_{\rm shell}} = \sum_k e_k - \widetilde E \; .
\label{eq02}
\end{equation}
Here the s.p.\ energies $e_k$ are simply the eigenvalues of a  mean-field
Hamiltonian with a Yukawa-folded s.p.\ potential at the desired  deformation.
The average energy $\widetilde E$ is evaluated using the Strutinsky prescription
\cite{Str66,NTS69} with an 6$^{\rm th}$ order correction polynomial. The pairing
correction is determined as the difference between the BCS energy \cite{BCS57}
and the single-particle energy sum and the average pairing energy \cite{NTS69}:
\begin{equation}
 E_{\rm pair} = E_{\rm BCS} - \sum_{k} e_k
            - \widetilde{E}_{\rm pair}\;.
\label{Epair}\end{equation}
In the BCS approximation the ground-state energy of a system with an even number
of particles and a monopole pairing force is given by
\begin{equation}
E_{\rm BCS} = \sum_{k>0} 2e_k v_k^2 - G(\sum_{k>0}u_kv_k)^2 - G\sum_{k>0} v_k^4
 -{\cal E}_0^\phi\;,
\label{EBCS}\end{equation}
where the sums run over the pairs of s.p.\ states contained in the pairing
window defined below.  The coefficients $v_k$ and $u_k=\sqrt{1-v_k^2}$ are the
BCS occupation amplitudes, and ${\cal E}_0^\phi$ is the pairing-energy
correction due to the particle number projection done in the GCM+GOA
approximation \cite{GPo86}
\begin{equation}
{\cal E}_0^\phi=\frac{\sum\limits_{k>0}[ (e_k-\lambda)(u_k^2-v_k^2)
        +2\Delta u_k v_k +Gv_k^4] / E_k^2}{\sum\limits_{k>0} E_k^{-2}}\;,
\label{Ephi}\end{equation}
where $E_k=\sqrt{(e_k-\lambda)^2+\Delta^2}$ are the quasi-particle energies, 
and $\Delta$ and $\lambda$ denote the pairing gap and the Fermi energy, 
respectively. The average projected pairing energy, for a pairing window of
width $2\Omega$, which is symmetric in energy with respect to the Fermi energy,
is equal to
\begin{equation}
\label{Epavr}
\begin{array}{l}
 \widetilde{E}_{\rm pair}=-\frac{1}{2}\,\tilde{g}\,\tilde{\Delta}^2+
 \frac{1}{2}\tilde{g}\,G\tilde{\Delta}\,
 {\rm arctan}\left(\frac{\Omega}{\tilde\Delta}\right)
  -\log\left(\frac{\Omega}{\tilde\Delta}\right)\tilde{\Delta}\\
 \hspace{2.0cm}
+\frac{3}{4}G\frac{\Omega/\tilde{\Delta}}{1+(\Omega/\tilde{\Delta})^2}/
  {\rm arctan}\left(\frac{\Omega}{\tilde{\Delta}}\right)-\frac{1}{4}G \; .
\end{array}
\end{equation}
Here $\tilde{g}$ is the average single-particle level density and 
$\tilde\Delta$ the average paring gap corresponding to a pairing strength $G$ 
\begin{equation}
 \tilde\Delta=2\Omega\exp\left(-\frac{1}{G\tilde{g}}\right) .
\label{Davr}\end{equation}
The width of the pairing window for protons or neutrons is chosen so as to
contain $2\sqrt{15\cal N}$ ($\cal N =\,$N or Z) s.p.\ states around the Fermi
surface. For such a window the pairing strength can be approximated by
\cite{PPS89}
\begin{equation} 
 G = \frac{g_0}{{\cal N}^{2/3} \, A^{1/3}} \; ,
\label{Gpair}\end{equation}
where $A$ is the mass number, and we have used the same value $g_0^p=g_0^n=
0.28\hbar\omega_0$ for protons and neutrons with $\hbar\omega_0= 41/A^{1/3}$
MeV. 

In the whole calculation the single-particle spectra are obtained by
diagonalization of a Hamiltonian with a Yukawa-folded mean-field potential 
\cite{DNi76,DPB16} having the same parameters as those used in Ref.~\cite{MNi95}.

The macroscopic-microscopic energy landscape of each nucleus is evaluated as
function of the different deformation degrees of freedom by using the Fourier
parametrization of the nuclear shape that we recently developed  \cite{PNB15}: 
\begin{equation}
\begin{array}{ll}
 \frac{\rho_s^2(z)}{R_0^2} =\! \sum\limits_{n=1}^\infty &\left[
 a_{2n} \cos\left(\frac{(2n-1) \pi}{2} \, \frac{z-z_{sh}}{z_0}\right)\right.\\
 &+\left. a_{2n+1} \sin\left(\frac{2 n \pi}{2} \, \frac{z-z_{sh}}{z_0}\right)
 \right]~,
\end{array}\end{equation} 
where $z_0$ is the half-length of the shape and $z_{\rm sh}$ locates the centre
of mass of the nucleus at the origin of the coordinate system.
It turns out that one can define what we will call ``{\it optimal
coordinates}'', $q_n$, through \cite{SPN17}
\begin{equation}
\left\{
\begin{array}{l}
 q_2=a_2^{(0)}/a_2 - a_2/a_2^{(0)} \\[1ex]
 q_3=a_3 \\[1ex]
 q_4=a_4+\sqrt{(q_2/9)^2+(a_4^{(0)})^2} \\[1ex]
 q_5=a_5-(q_2-2)a_3/10  \\[1ex]
 q_6=a_6-\sqrt{(q_2/100)^2+(a_6^{(0)})^2} \;\;,
\end{array} \right.
\label{qi}\end{equation}
in such a way that the liquid-drop energy as function of the elongation $q_2$ 
becomes minimal along a track that defines the liquid-drop path to fission. The
$a^{(0)}_{2n}$ in Eq.\ (\ref{qi}) are the expansion coefficients of a spherical
shape given by
$a^{(0)}_{2n}=(-1)^{n-1}\frac{32}{\pi^3\,(2n-1)^3}$. 
Non-axial shapes can easily be obtained assuming that, for a given value of the
$z$-coordinate, the surface cross-section has the form of an ellipse with
half-axes $a(z)$ and $b(z)$ \cite{SPN17}:
\begin{equation}
   \hspace{-0.3cm}  \varrho_s^2(z,\varphi) = \rho^2_s(z) 
                \frac{1-\eta^2}{1+\eta^2+2\eta\cos(2\varphi)}
   \hspace{0.3cm} \mbox{with} \hspace{0.3cm} 
 \eta = \frac{b-a}{a+b}
\label{eta}\end{equation}
where the parameter $\eta$ describes the non-axial deformation of the nuclear
shapes. The volume conservation condition requires that
$\rho_s^2(z)=a(z)b(z)$.

A similar kind of investigation has been undertaken in Ref.~\cite{MSB12} for a
much wider range of nuclei (7206 isotopes with masses between A$\,=\,$31  and
A$\,=\,$290) but for a much more restricted deformation space (3 deformation 
parameters $\epsilon_2,\, \epsilon_4$ and $\gamma$ corresponding respectively 
to quadrupole, hexadecapole and axial asymmetry), whereas  our deformation 
space contains the $q_n$ deformation parameters of Eq.\ \ref{qi}) plus one 
axial asymmetry parameter $\eta$ defined in Eq.\ (\ref{eta}). The study of 
Ref.\ \cite{MSB12} was also carried out using the  macroscopic-microscopic
method, but with the finite-range liquid droplet model (FRLD), while in the 
present study we use the Lublin-Strasbourg Drop model (LSD) \cite{PDu03}. 
As it was shown in Ref.~\cite{PDu03}, the use of deformations of higher 
multipolarity, that are different in both models, induces a different 
stiffness of the PES, what has a non negligible effect on the energy landscape.
At difference from Ref.\ \cite{MSB12} where a pairing treatment in the BCS 
approach was used together with the Lipkin-Nogami prescription to take into 
account the effect of the projection on the correct particle number, we use 
an approximate GCM+GOA particle-number projection in our approach 
\cite{GPo86,PPS89}, which yields results that are much closer to the  exact 
ones, in particular in the weak-pairing limit.
\begin{figure}[ht]
\includegraphics[width=0.95\columnwidth]{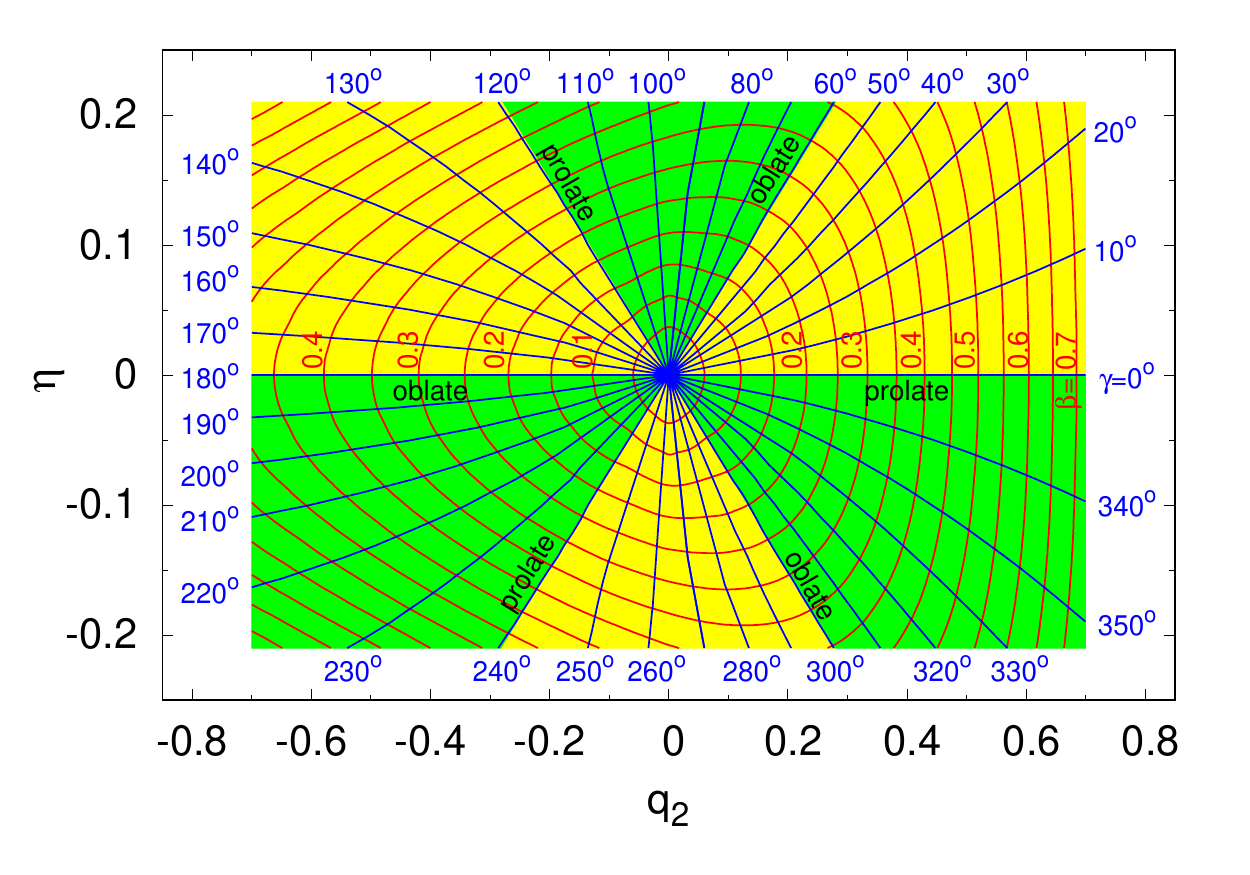}
\vspace{-3mm}
\caption{Relation, for a spheroidal shape, between the elongation $q_2$ and 
non-axiality $\eta$ parameters \cite{SPN17} on one side, and the traditional 
Bohr $\beta$ and $\gamma$ deformations \cite{Boh52,KSP76} on the other side.}
\label{be-ga} 
\end{figure}

The above description of non-axial shapes of deformed nuclei ($q_2,\eta$) is
more general than the commonly used ($\beta,\,\gamma$) parametrization by  Bohr
\cite{Boh52,KSP76}. For spheroidal shapes both descriptions  are, however,
equivalent.  As one can see from Fig.~\ref{be-ga}, where the two
parametrizations are  compared, the periodicity of the nuclear shapes by an
angle of 60$^{\circ}$  is similar in the ($q_2,\eta$) and in the
($\beta,\,\gamma$) planes.  One has, however, to keep in mind that this
regularity is partially spoiled when higher multipolarity deformations $q_n$
($n>2$) come into play,  which makes that our ($\eta,\,q_2,\,q_3,\,q_4,\,q_6$)
shape parametrization  is substantially more general than the 3-dimensional 
$(\varepsilon_2, \varepsilon_4,\gamma)$ parametrization used in
Ref.~\cite{MSB12}. It is only for the very special case  of spheroidal shapes
that both parametrization coincide.

It is important, in this context, to stress that there are different points in
the ($\beta,\,\gamma$), as in the ($q_2,\eta$) plane, that correspond, when
higher $q_n, \; n>2$ degrees of freedom are neglected (and only under that
assumption!), to precisely the \underline{same} shape, with the only difference
that the  axes of the coordinate system have been interchanged (like $y
\leftrightarrow  z$). As an example, the point ($\beta\!=\!0.4,\,\gamma\!=\!0$),
described by  ($q_2\!=\!0.42,\,\eta\!=\!0$) in the new parametrization,
identifies obviously  the same shape as ($\beta\!=\!0.4,\, \gamma \!=\!
60^{\circ}$) or equivalently in the new parametrization by ($q_2\!=\!-0.21,\,
\eta\!=\!0.16$). When investigating potential energy landscapes, where the
triaxial degree of freedom is taken  into account, one therefore should be
extremely careful, not to consider as  two different configurations points in
the ($q_2,\eta$) deformation plane  that are nothing but $\gamma\!=\!
60^{\circ}$ {\it rotation images} of one  another.

\section{Results}
\label{Res-sec}

The calculations of the potential-energy surfaces were performed for the
isotopic chains of even-even Pt, Hg and Pb nuclei, with the same range of
neutron numbers between $N$=92 and $N$=126. The macroscopic-microscopic energy
of each isotope was evaluated in a four-dimensional (4D) deformation space
spanned by the $q_2$, $q_3$, $q_4$, and $\eta$ coordinates [see Eqs. (\ref{qi})
and (\ref{eta})]. In parallel, a  calculation made in the 3D space of the $q_2$,
$q_4$, and $q_6$ deformation  parameters was carried out in order to test the
effect of higher-order  deformations on the PES in the considered nuclei. 
\begin{figure}[ht]
\vspace{-3mm}
\includegraphics[width=0.95\columnwidth]{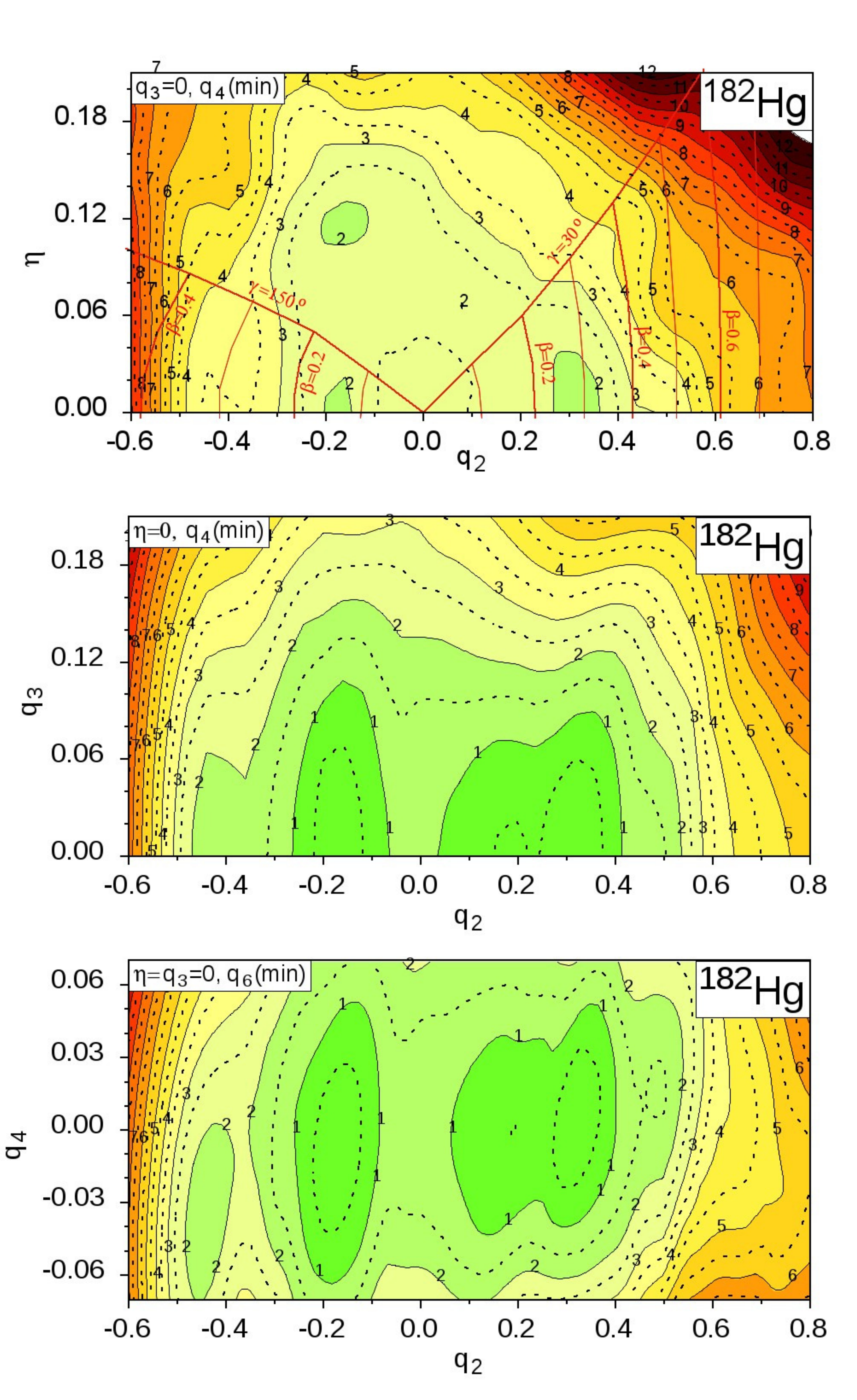}
\vspace{-2mm}
\caption{Potential energy landscape of $^{182}$Hg on the ($q_2,\,\eta$)
(top),($q_2,\,q_3$) (middle) and ($q_2,\,q_4$) (bottom) planes. The
($\beta,\gamma$) grid (red lines) is shown in the top map.}
\label{Hg182-maps}
\end{figure}
Our calculations in the 4D deformation space spanned by the $q_2$, $q_3$, $q_4$,
and $\eta$ coordinates, clearly show that the left-right asymmetry degree of 
freedom does not decrease the potential energy at any of the studied 
deformations, in any of the here considered nuclei, as one can see in the
($q_2,\,\eta$), ($q_2,\,q_3$) and ($q_2,\,q_4$) PES presented in Fig.\
\ref{Hg182-maps} for the $^{182}$Hg nucleus, chosen here as a representative 
case for all the nuclei in the present study. The PES shown are generally 
defined as the difference of the total nuclear energy (\ref{eq01}) of the 
deformed nucleus and the one of the corresponding spherical shape evaluated 
using the LSD macroscopic model. All energy points in the ($q_2,\,\eta$) map 
(top of Fig.\ \ref{Hg182-maps}) are minimized with respect  to the $q_3$ and
$q_4$ deformation degrees of freedom. The additional red lines  indicate the
corresponding ($\beta,\, \gamma$) coordinates, with  two sectors, namely $0
\!\leq\! \gamma \!\leq\! 30^o$ and  $150^o \!\leq\! \gamma \!\leq\! 180^o$
explicitly indicated, which  correspond to the lower and upper part of
the ($\beta,\,\gamma$) map  in which only the $0\leq\gamma\leq 60^o$ sector is 
traditionally displayed (see e.g.\ Ref.~\cite{KSP76}).  In addition, red lines
representing $\beta = 0.1,\, 0.2,\, \dots$ values are  drawn.  The part of the
($q_2,\,\eta$) map between the $\gamma=30^o$ and $\gamma=150^o$  lines is
usually omitted when one projects a multidimensional  macro-micro or
self-consistent PES onto the ($\beta,\,\gamma$) plane, loosing hereby, as we
have explained above, some information on the impact  of higher multipolarities.
We have therefore decided to show in the following the full ($q_2,\,\eta$) maps
in order not to lose part of the results, which could be important when e.g.\
the deepest minimum would appear in this upper part of the  map.

In the middle panel of Fig.~\ref{Hg182-maps} is shown the PES in the
$(q_2,q_3)$ plane which has been obtained for the axial symmetric case
($\eta\!=\!0$) after minimization with respect to the $q_4$ degree of freedom.
In this and in the rest of the energy landscapes shown in the following, solid
lines correspond to layers separated by 1 MeV, while the distance between dashed
and neighbouring solid lines is of 0.5 MeV. The labels on the layers denote
energies in MeV.  
\begin{figure}[ht]
\includegraphics[width=0.85\columnwidth]{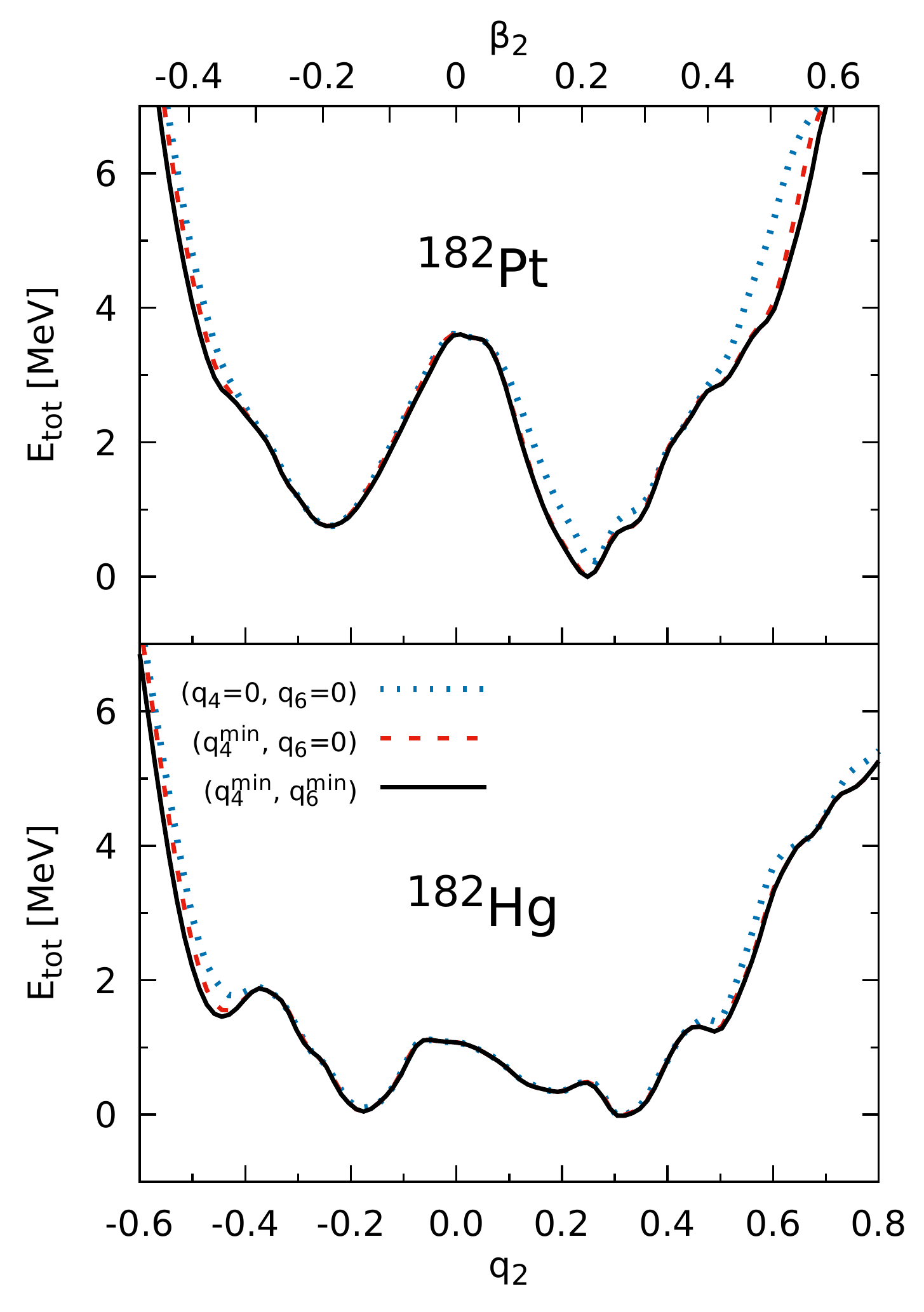} \vspace{-2mm}
\caption{Potential energy of $^{182}$Pt (top) and $^{182}$Hg (bottom) as 
functions of the elongation parameter $q_2$, with the corresponding $\beta_2$ 
values on a scale on top of the figure. Different line types indicate 
minimization  of the energy with respect to different shape parameters: $q_4$
and $q_6$ set to zero (blue dotted line), minimization with respect to $q_4$,
but $q_6$ set to zero (red dashed line), minimization with respect to $q_4$ 
and $q_6$ (black solid line).} 
\label{Pt-Hg182} 
\end{figure}

The PES displayed in the bottom panel of Fig.~\ref{Hg182-maps} shows the  
energy landscape in the ($q_2,\,q_4$) plane. The structure of the PES is seen 
to be rather complex and quite a few local minima can be observed in this
cross-section of our 5D deformation space, two on the oblate ($q_2\le 0$) and
three on the prolate side. One has to bear in mind, however, that one is looking
here only at a 2D cross-section of a 5D deformation space and that the stability
of these local minima with respect to the other deformation degrees of freedom,
especially the non-axial $\eta$, should be carefully studied in each case. Also
the role of the higher multipole deformation $q_6$ needs to be analysed. We have
therefore tried to investigate the impact of these higher-multipole degrees of
freedom on the deformation energy of $^{182}$Pt and $^{182}$Hg. The results of
this study is shown in Fig.~\ref{Pt-Hg182}, where the energy as function of the
$q_2$ elongation parameter is shown in a 1D cross-section of the PES and where
different lines indicate the importance of different higher-order deformations.
These deformation energies which are typical for all the considered nuclei in
the three here investigated isotopic chains, clearly show that the role of $q_6$
deformations is practically negligible, thus  demonstrating that the ``{\it
optimal coordinates}'' $q_n$, defined in  Eq.~(\ref{qi}), have, indeed, been
skilfully chosen.

\begin{figure*}[ht!]
\includegraphics[width=2.10\columnwidth]{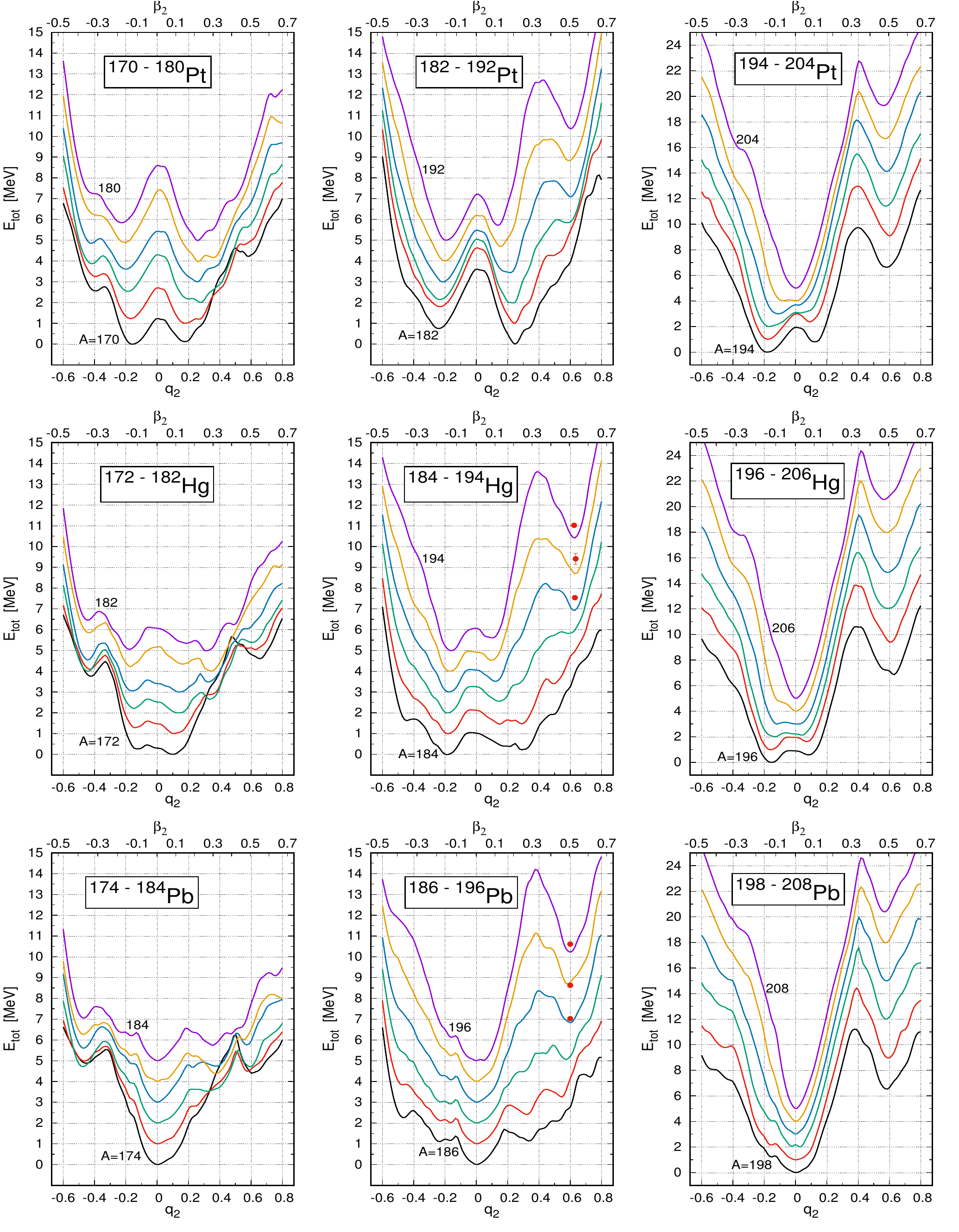} 
\caption{Potential energies of Pt, Hg, and Pb isotopes with 92$\le$ N $\le$ 126
as a function of the elongation parameter $q_2$ or quadrupole deformation
$\beta_2$ (upper axis). Energies are minimized with respect to $q_4$ and
$q_6$ parameters. All local minima correspond to reflection symmetric
shapes ($q_3=0$). Experimental energies of SD minima (red dots) are taken
from Ref.~\cite{LLL16}.} 
\label{Pt-Hg-Pb}\end{figure*}

An overview of the deformation energies as function of the elongation parameter
$q_2$ is shown in Fig.\ \ref{Pt-Hg-Pb} for the nuclei of the three Pt, Hg, and
Pb isotopic chains with neutron numbers $92 \leq N \leq 126$, where the
potential energy of each isotope is minimized with respect to $q_4$ and $q_6$.
The left-right asymmetry degree of freedom $q_3$ is not included in this study,
since for all the here considered nuclei it does not play any noticeable role as
demonstrated above.  Lines corresponding to different isotopes are drawn in
different colours and the ground-state energy minimum of each isotope is shifted
by 1 MeV with respect to the previous isotope in order to separate the curves.
In each diagram the bottom and top lines are marked by the corresponding mass
number A for better visibility. The experimental excitation energies of the 
super-deformed (SD) minima are also shown when available, as in the case of Hg 
and Pb nuclei.

In the upper parts of each diagram in Fig.\ \ref{Pt-Hg-Pb} the corresponding 
values of the quadrupole deformation $\beta_2$ are also given. This deformation
parameter is defined by the harmonic expansion of the radius of a deformed
nucleus:
\begin{equation}
R(\theta)=R_0(\{\beta_\lambda\})\left(1+\sum\limits_{\lambda=1}^{\lambda_{max}}
         \beta_\lambda Y_{\lambda 0}(\theta,0)\right)\,
\label{shexp}\end{equation}
where $Y_{\lambda\mu}(\theta,\varphi)$ are the spherical harmonics. Such an
expansion is frequently used in publications dealing with PES calculated in  the
mac-mic approach (see e.g.\ Ref.~\cite{JKS17}), 
\clearpage

\noindent
but turns out to be very  slowly converging at large nuclear  elongations (as 
shown in Refs.~\cite{PNB15,SPN17,DPB75}).
The experimental energies corresponding to the bottom of the SD bands, measured
with respect the ground state, are marked by the red points in Fig.\
\ref{Pt-Hg-Pb}. These experimental data are taken from Ref.~(\cite{LLL16},
Fig.~46) and found  in all cases to be in quite good agreement with our
theoretical predictions.

Quite pronounced oblate and prolate minima around $q_2 \!\approx\! 0.2$ and $q_2
\!\approx\! -0.2$ are found in all Pt isotopes, except that one finds,  when
looking just at such a kind of 1D deformation energy, that beyond $^{194}$Pt
(with neutron number larger than  N$\approx$116) the prolate minimum at $q_2
\!\approx\! 0.2$ gradually looses its importance at the profit of a
super-deformed (SD) shape isomer located around $q_2 \!\approx\! 0.6$. One has,
however, to keep in mind that we have, for the moment, completely left out of
our considerations the non-axial degree of freedom. As we will see in Fig.\
\ref{Pt-e21} in Subsection \ref{Pt-sec}, that analysis will change, when we will
take that additional degree of freedom into account. While in $^{170}$Pt the
ground-state deformation is oblate, situated only about 200 keV below the
prolate minimum, heavier Pt isotopes with $172\leq A\leq 186$ show a prolate
ground state. The situation changes again in the case of still heavier Pt
isotopes $188\leq A\leq 200$, where the oblate minimum becomes again deeper with
a deformation energy, however, that gets smaller and smaller with growing
neutron number as one approaches to the magic number $N=126$ when the isotopes
$^{202-204}$Pt become spherical. The largest deformation energy of $E_{\rm
def}=E_{\rm tot}^{\rm g.s.}-E_{\rm tot}^{\rm sph.} \approx 3.5$ MeV is predicted
by our calculation to occur in $^{178-184}$Pt. A delicate structure appears on
the prolate side in $^{172-178}$Pt isotopes when two minima around $q_2=0.25$
and $q_2\approx 0.35$ begin to compete. At this point one cannot really judge
about the stability of these local minima and an eventual shape coexistence by
looking at such a 1D plot only. A more extended study which takes also the
non-axial degree of freedom into account is necessary and will be carried out in
Subsection 3.1.

Apart from the minima at deformations typical for the ground state, strongly 
oblate minima at $q_2 \approx -0.4$ are found in $^{170-176}$Pt isotopes, 
minima that are separated from the normal-deformed oblate one by a small 
barrier of only $\approx$ 0.5 MeV. In addition, a rather strongly deformed 
pronounced prolate minimum at $q_2 \approx 0.6$ is observed both in the 
lightest here studied Pt isotope $^{170}$Pt, as well as in Pt isotopes with 
A$\geq\,$186, with a barrier which separates them from the normal-deformed 
prolate minimum and thus guarantees their stability, a barrier that gradually
grows with increasing neutron number to reach about 3 MeV in $^{204}$Pt. 
In most cases these SD isomeric states are, however, rather high in energy 
above the ground state ($E_{\rm tot}^{\rm isomer}- E_{\rm tot}^{\rm g.s.}
\approx 6$ MeV), and will therefore be difficult to be observed experimentally.
On the other hand, the moment of inertia of nuclei in such SD bands is several 
times larger as compared to a typical deformed ground-state band \cite{SBP73}, 
which has a  significant influence on the high-spin level structure since the SD 
states can  appear at energies close to or even lower than the rotational states 
built on  the ground-state.

The situation in the Hg isotopes (middle row of Fig.\ \ref{Pt-Hg-Pb}) is  even
more complex. Apart from the strongly deformed oblate minimum at $q_2 
\!\approx\! -0.45$ observed in the lighter isotopes $^{172-184}$Hg,  separated
from the ground state minimum by a barrier of up to 1 MeV, a very  elongated
shape isomer is present at $q_2 \!\approx\! 0.6$ in $^{172-176}$Hg,  disappears
beyond $^{176}$Hg, and reappears again in the $^{190-206}$Hg  isotopes. This
strongly deformed shape isomer is {\it protected} from a decay  towards to the
normally deformed states by a fairly high barrier and is located between 4 MeV
(in $^{190}$Hg) and 16 MeV (in $^{206}$Hg) above the ground state. In
$^{176-186}$Hg additional low-lying prolate or oblate minima appear, thus giving
the chance to observe a shape-coexistence in these nuclei. One has of course to
study the stability of these local minima with respect to non-axial
deformations, an investigation that will be carried out in Subsection
\ref{Hg-sec}. 

The ground state of all considered $^{174-208}$Pb isotopes is found to be 
spherically symmetric, but in all of them deformed local minima can be 
observed. Similar as in the case of the Mercury isotopes, strongly deformed 
oblate minima with $q_2 \!\approx\! -0.45$ appear in the isotopes with $174 
\!\leq A\! \leq 186$, located again rather high above the ground state, at 
about 5 MeV in $^{174}$Pb and 2.5 MeV in $^{186}$Pb. Very deformed prolate 
configurations with a main-axis ratio close to 2 ($q_2 \approx 0.6$) are 
predicted by our calculations in the five lightest isotopes $^{174-182}$Pb and 
in isotopes with $N \gtrsim 104$. It is interesting to notice that some 
additional local minima appear at smaller deformations $q_2 \!\approx\! -0.2$ 
and $0.2 \!\lesssim\! q_2 \lesssim 0.5$ in $^{182-188}$Pb isotopes, where even 
several oblate and prolate deformed minima are predicted. These results, 
already reported in our previous study \cite{NPB17}, are in line with the 
experimental observations (see e.g.\ Ref.~\cite{And00}). The stability of  these
minima with respect to the non-axial degree of freedom will be discussed in the
subsequent Subsection \ref{Pb-sec}. 

In the three lightest Pb isotopes (A=174 to 178) one observes a sudden  wobbling
of the energy at $q_2\approx 0.5$ and the appearance of a local  SD minimum at
$q_2\approx 0.6$. These are induced by the minimization process, when i.e.\ the
energy minimum jumps with increasing elongation $q_2$ from one  to another
valley in the ($q_4,q_6$) plane. 

\begin{figure}[ht!]
\includegraphics[width=1.0\columnwidth]{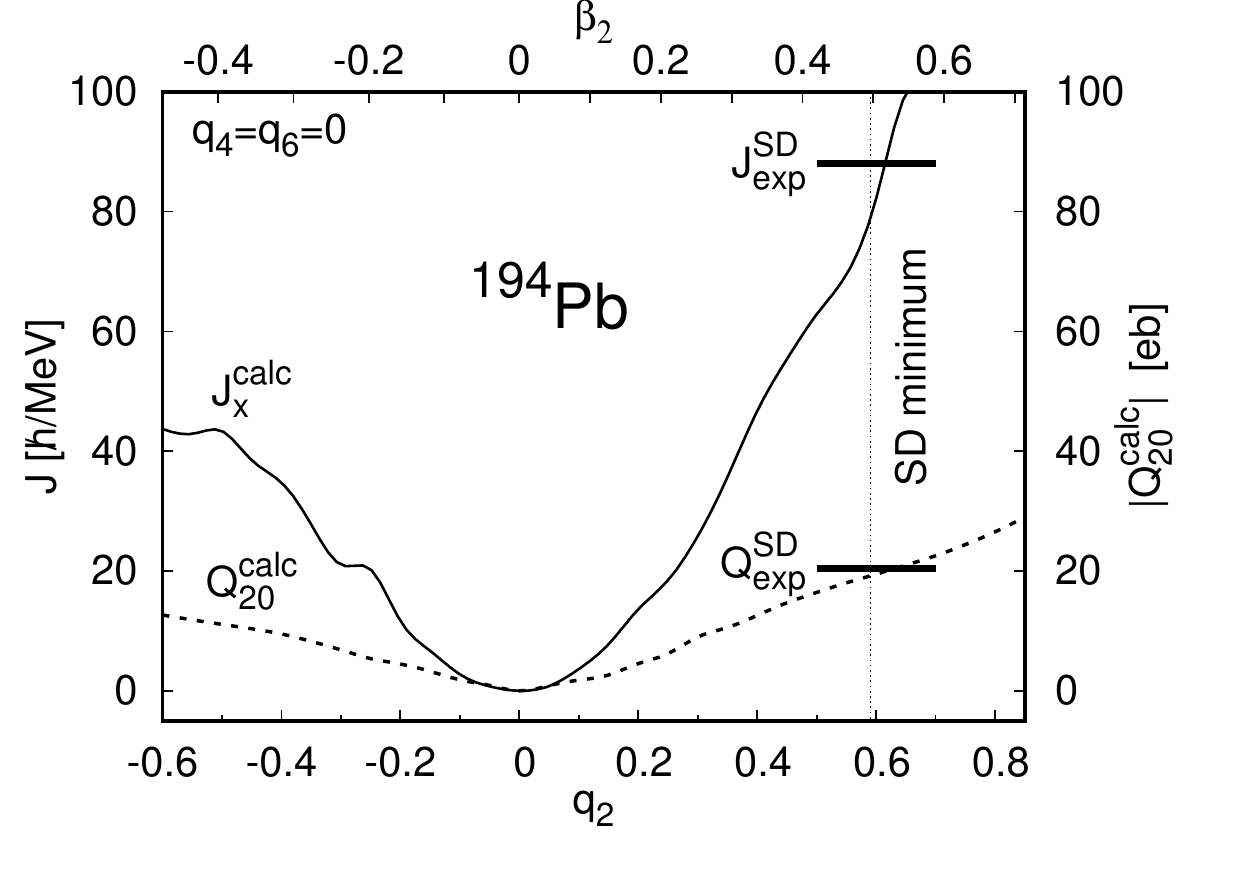}
\caption{Cranking moment of inertia (solid line) and electric quadrupole moment
(dashed line) of $^{194}$Pb as a function of the elongation $q_2$ (bottom scale) 
or $\beta_2$
deformation (top scale). The experimental values of the quadrupole moment and
the moment of inertia extracted from the SD rotational band are marked by the
thick horizontal lines. The data are taken from Ref.~\cite{SZF02}.Thin dotted
vertical line corresponds to the predicted equilibrium deformation of the SD
isomer.}
\label{JQ}
\end{figure}
\begin{figure}[ht!]
\includegraphics[width=1.0\columnwidth]{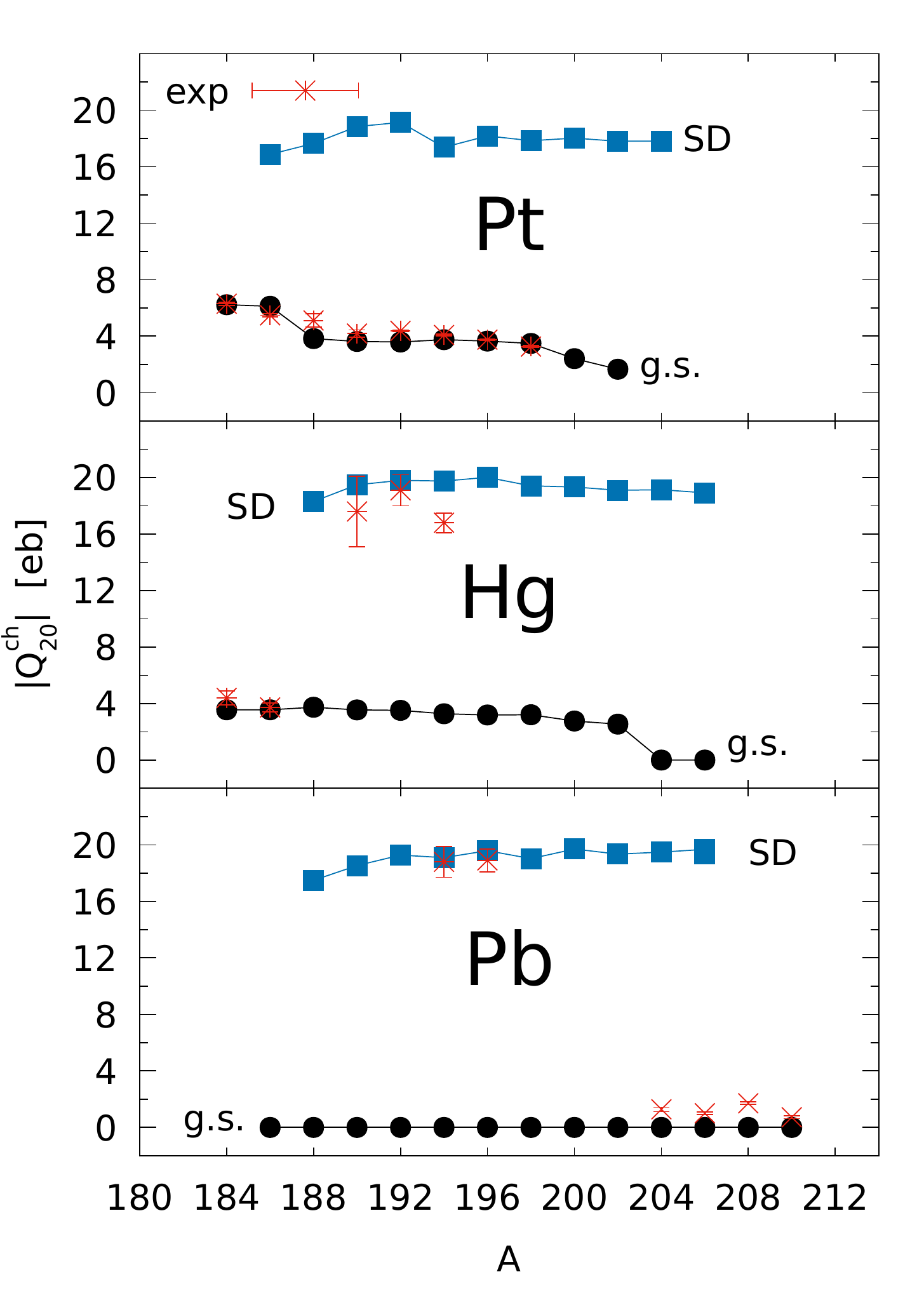}
\caption{Predicted electric quadrupole moments of Pt, Hg, and Pb isotopes in
the  ground-state (black points) and in the super-deformed isomeric 
states (blue squares) are compared with the experimental data (stars) taken
from  Refs.~\cite{LLL16,SZF02,NUDAT}.}
\label{QQ}
\end{figure}

One could ask the question about the accuracy of our theoretical predictions.  A
comparison with available data, taken from Refs.~\cite{LLL16,SZF02,NUDAT},  is
presented in Figs.~\ref{JQ} and \ref{QQ}, where Fig.~\ref{JQ} shows the  moment
of inertia $J_{\rm x}^{}$ and the charge quadrupole moment  $Q_{20}^{}$ as
function of the elongation parameter for $^{194}$Pb, and  Fig.~\ref{QQ} the
charge quadrupole moment in the ground state and the SD shape  isomeric state
for the isotopic chains of Pt, Hg and Pb.   The experimental value $J_{\rm
x}^{\rm exp}$ of the moment of inertia of the  SD shape isomer in Fig.\ \ref{JQ}
is evaluated from the energies of two lowest  experimentally observed members
($4^+$ and $6^+$) of the SD rotational band  \cite{SZF02}.  One can see that the
experimental values ($J_{\rm x}^{\rm exp}$ and  $Q_{\rm 20}^{\rm exp}$), given
by the horizontal lines in Fig.\ \ref{JQ}  coincide quite well, at the predicted
deformation of the SD minimum (vertical  dotted line), with the theoretical
values for both these quantities. Another measure of the accuracy of our
predictions is presented in Fig.\  \ref{QQ}, where the theoretical absolute 
values of the charge quadrupole moment for  the ground state (black points) and 
the SD minimum (blue squares) of Pt, Hg,  and Pb are compared with the 
corresponding experimental data (red stars) taken  from Refs.~
\cite{LLL16,SZF02,NUDAT}.

We would like to stress at this point that all results, obtained in models 
taking only axial symmetric shapes into account and giving an indication of a 
possible existence of shape isomers or shape coexistence, cannot seriously be 
considered as reliable without checking the stability of the identified minima
with respect to other degrees of freedom, among which the non-axial deformation
seems to be the most important one. That is precisely what will be investigated
in the subsequent subsections.


\subsection{Pt nuclei}
\label{Pt-sec}

\begin{figure*}[t!]
\includegraphics[width=2.0\columnwidth]{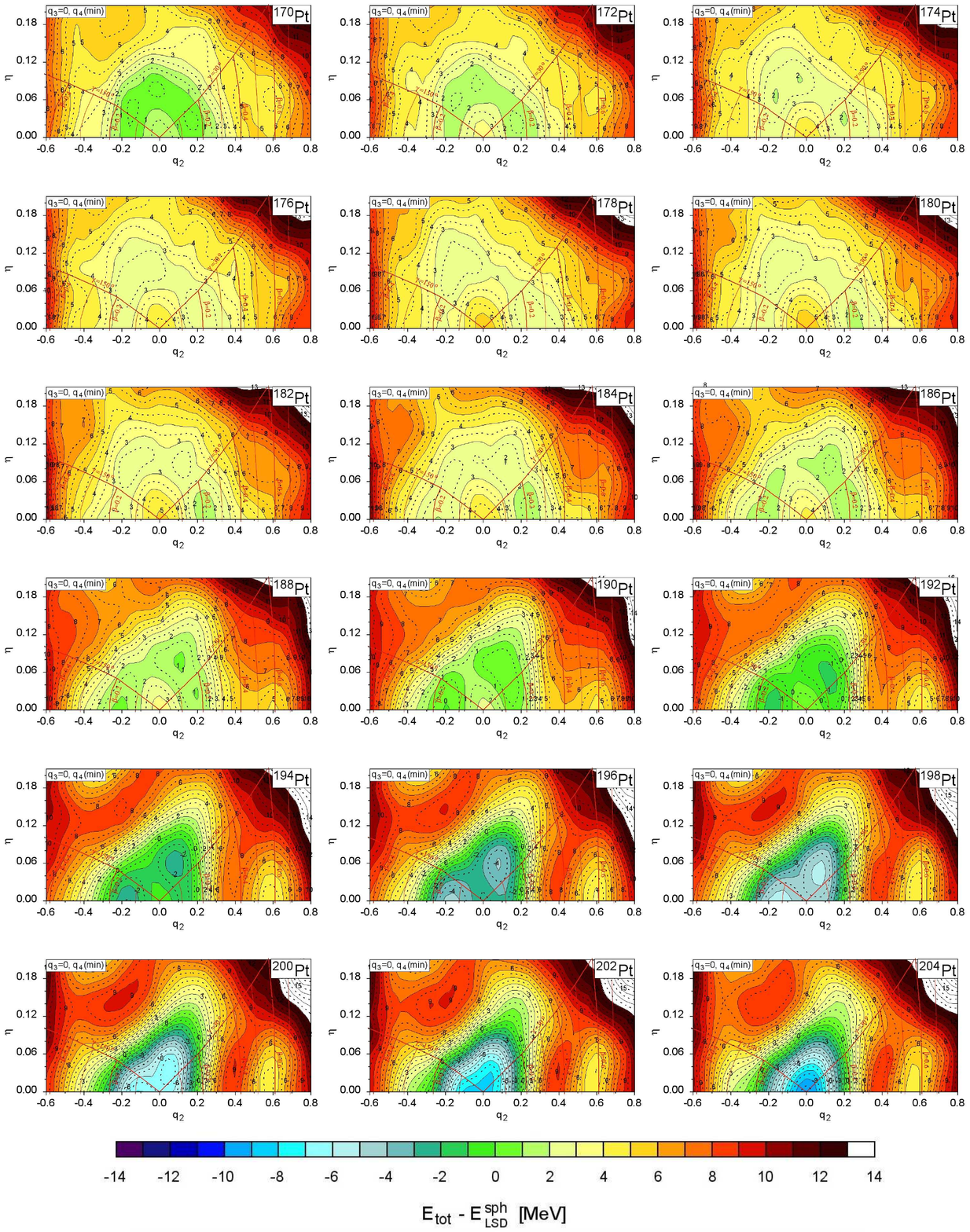}
\vspace{-5mm}
\caption{Potential energy landscape of eighteen even-even nuclei of the Pt 
isotopic chain on the ($q_2,\,\eta$) plane. Each point is minimized with 
respect to $q_4$ deformations. Solid lines correspond to the layers separated 
by 1 MeV, with dashed and solid lines separated by 0.5 MeV. Labels at the 
layers denote energies in MeV.}
\label{Pt-e21}
\end{figure*}

The potential energy surfaces of $^{170-204}$Pt isotopes on the ($q_2,\,\eta$)
plane are displayed in Fig.~\ref{Pt-e21}.  The energy in each of the deformation
points on these maps  is minimized with respect to $q_4$ deformations only,
since we have seen, as explained above, that $q_3$ and $q_6$ degrees of freedom
are practically  ineffective in these nuclei at the investigated deformations. A
very elongated local minimum appears  here at $q_2 \!\approx\! 0.6$ in the two
lightest $^{170-172}$Pt isotopes, when a non-axial deformation is taken into
account. It turns out that this local minimum is deeper here than in Fig.\
\ref{Pt-Hg-Pb} where that degree of  freedom is absent. Both these isotopes also
appear to be unstable in their  ground state ($|q_2|\approx 0.2$) with respect
to the non-axial degree of  freedom. They can therefore be considered as good
examples of the so-called  $\gamma$-instability. Please recall in this respect
the discussion related to  the two different descriptions, ($q_2,\,\eta$) and
($\beta,\,\gamma$), of the  non-axial degree of freedom at the end of Section
\ref{Sect.02} and  visualized in Fig.~\ref{be-ga}. 
%
\begin{figure*}[t!] 
\includegraphics[width=2.0\columnwidth]{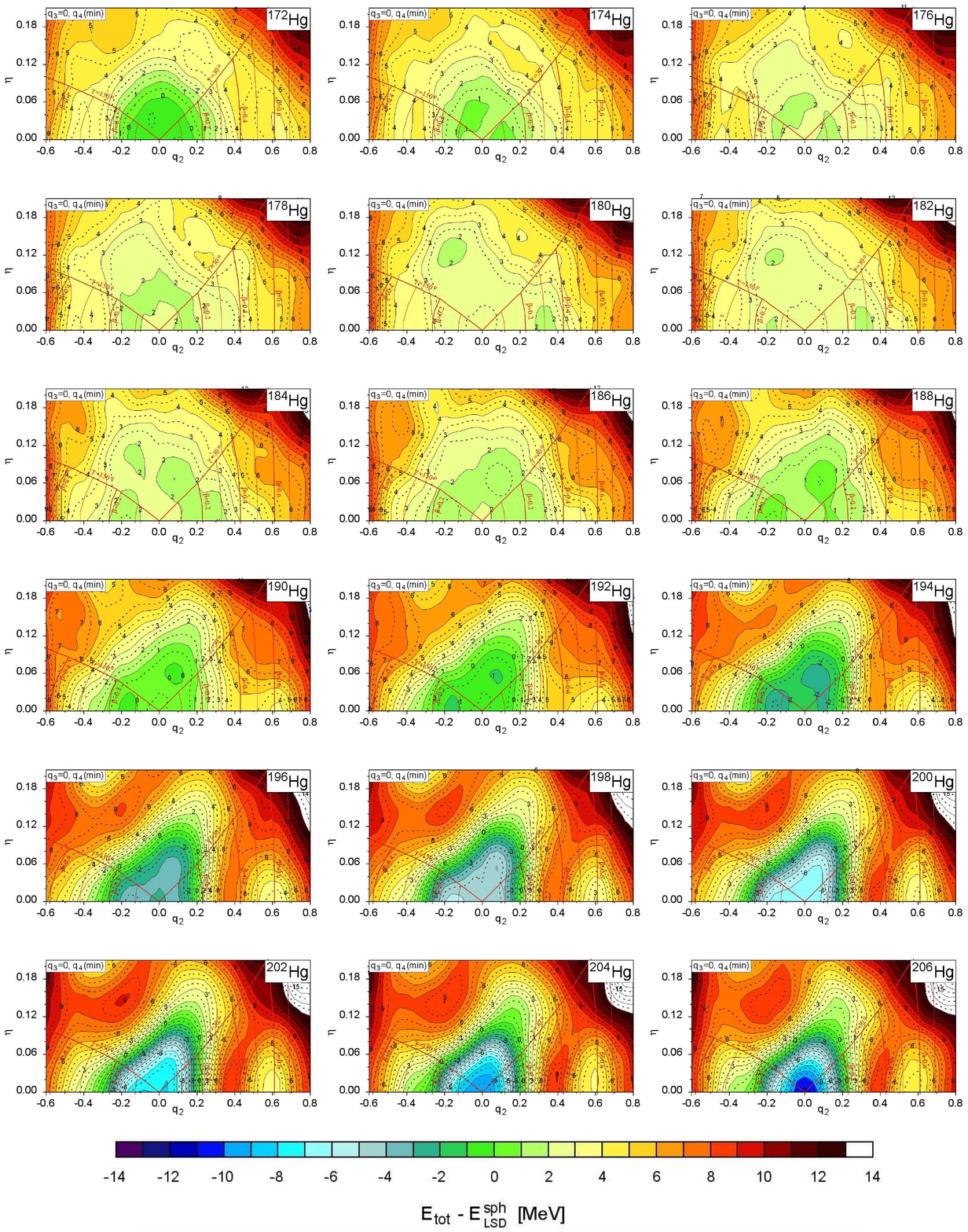}
\vspace{-5mm} 
\caption{The same as in Fig.~\ref{Pt-e21} but for the Hg isotopes.}
\label{Hg-e21} 
\end{figure*}
%
\begin{figure*}[t!]
\includegraphics[width=2.0\columnwidth]{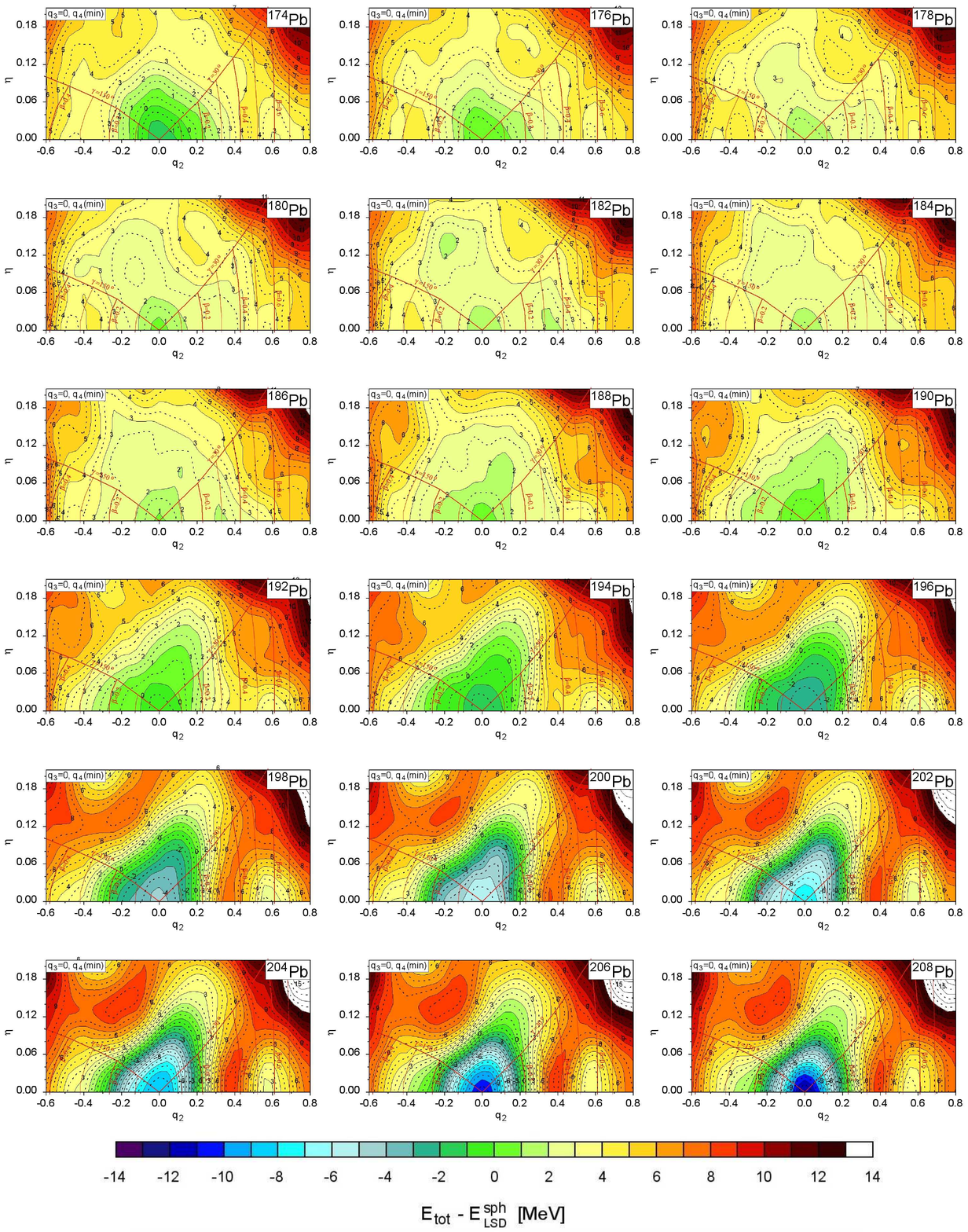}
\vspace{-5mm}
\caption{The same as in Fig.~\ref{Pt-e21} but for the Pb isotopes.}
\label{Pb-e21}
\end{figure*}
\noindent
When looking at the deformation energies of the $^{174-180}$Pt nuclei in Fig.\
\ref{Pt-Hg-Pb}, where axial symmetry had been imposed, one has the impression
that there is a coexistence of a prolate  minimum at $q_2 \approx 0.28$ with an
oblate minimum at $q_2 \approx -0.20$  separated by a barrier of about 1.5 to
2.5 MeV. When, however, taking the non-axial degree of freedom into account,
as this has been done in the  calculations represented in Fig.\ \ref{Pt-e21},
the picture changes. When  looking e.g.\ at the case of $^{178}$Pt, one again
finds a coexistence of two  minima, one prolate at ($q_2=0.27,\, \eta=0$), the
other triaxial at  ($q_2=-0.15,\, \eta=0.09$), which are separated by a much
lower barrier now, that now the non-axial degree of freedom is taken into
account Remembering  however the discussion at the end of Section \ref{Sect.02},
one concludes that  these two local minima just represent the \underline{same}
prolate shape, since  the triaxial minimum with negative $q_2$ value is
nothing but the  {\it symmetry image} of the axially symmetric prolate
ground-state minimum,  as can easily be  verified by taking into account the
60$^{\circ}$ $\gamma$ symmetry represented  in Fig.\ \ref{be-ga}. The same
situation is observed in all of the  $^{174-180}$Pt isotopes.  Let us recall at
this point that the above sextant symmetry, which is exact in  the case of
spheroid shapes,  is broken when higher multipolarity deformations are taken
into account. In Ref.~\cite{MSB12} the special choice of the axial and
non-axial  hexadecapole deformations, originally proposed by Rohozi\'nski and
Sobiczewski \cite{RSo81}, was used to keep this 60$^{\circ}$ regularity. The
shape  parametrization used in the present paper does not posses exactly this
symmetry. One could in principle introduce similar constraints as in
Ref.~\cite{RSo81}  on the $q_i$ deformation parameters to restore such a sextant
symmetry, but such constraints would cause a loss of generality of our
deformation space which we  want to avoid. A chance for a oblate-prolate shape
coexistence  appears, however, in the $^{182-186}$Pt isotopes with a small
barrier between  prolate and oblate minima. Also in $^{188}$Pt two minima are
observed, one  oblate at ($q_2 \!\approx\! -0.2,\, \eta =0$) and another one
triaxial at ($q_2  \!\approx\! 0.2,\,\eta \!\approx\! 0.02$). In the heavier Pt
isotopes with $190 \!\leq\! A \!\leq\! 200$, oblate shapes are preferred in the
ground state, whereas, approaching the N$\,=\,$126 shell closure, the
$^{202-204}$Pt nuclei turn out to be spherical. The very deformed shape isomers
around $q_2 \approx 0.6$ observed for the $^{188-204}$Pt isotopes in Fig.\
\ref{Pt-Hg-Pb} persist when non-axial deformations are taken into account,  with
a slight reduction (of up to 0.5 MeV) of the barrier height between these
isomers and the ground state for the four heaviest isotopes.


\subsection{Hg nuclei}
\label{Hg-sec}

Just in the same way as done in Fig.\ \ref{Pt-e21} for the Pt isotopes, we show 
in Fig.~\ref{Hg-e21} the deformation energies of the Mercury isotopes 
$^{172-206}$Hg on the ($q_2,\,\eta$) plane.  One observes that the shape
isomeric states corresponding to both the very  oblate deformations ($q_2
\!\approx\! -0.45$) in $^{172-182}$Hg,  and the very prolate deformations ($q_2
\!\approx\! 0.6$) in  $^{172-176}$Hg and in $^{190-206}$Hg all survive when the
non-axial degree of  freedom is taken into account with the exception  of the
prolate isomer in $^{174}$Hg and $^{176}$Hg which becomes slightly triaxial with
$\eta \approx 0.03$. A significant reduction of the height of the barrier
between the shape-isomeric states with $q_2 \approx 0.6$ and the ground state
minimum can, however, be observed in the $^{200-206}$Hg isotopes due to  the
non-axial degree of freedom. The ground state of $^{172}$Hg and of
$^{200-206}$Hg are spherically symmetric, while in the $^{174-176}$Hg  isotopes
it is the prolate minimum that corresponds to the ground state.  The additional
triaxial minima observed at ($q_2 \!\approx\! -0.05,\, \!\eta\! \approx 0.03$)
correspond again to the 60$^\circ$ mirror image of that ground-state
deformation, as explained at the end of Section \ref{Sect.02}. For the
$^{178-188}$Hg isotopes there should be a chance of a shape-coexistence at small
deformations, while for isotopes between $^{190}$Hg and $^{198}$Hg the
ground-state minimum corresponds to an oblate shape.


\subsection{Pb nuclei}
\label{Pb-sec}

The $^{174-208}$Pb isotopes are analysed in the same way in Fig.\ \ref{Pb-e21}
through PES landscapes in the ($q_2,\,\eta$) plane. The situation turns out,
however, to be slightly different here, since the ground state is found to be
spherically symmetric in all the considered isotopes and no competition between
prolate and oblate ground-state deformation is present. The issue of a possible
shape-coexistence can, in fact, not be limited to prolate versus oblate shapes,
but is much more general, as can be seen through the ($q_2,\,\eta$) maps at
small deformations in the $^{178-186}$Pb isotopes, where the landscapes in
$\eta$ directions are very soft, thus allowing for local minima, particularly
visible in the maps of $^{182-184}$Pb. Similarly as in the case of the Pt
and Hg isotopes, a well deformed prolate shape isomer at $q_2 \approx 0.6$ is
predicted in practically all Pb isotopes except $^{182-188}$Pb. This shape
isomer is particularly visible in the heavier isotopes, as already noticed on
the lower part of Fig.\ \ref{Pt-Hg-Pb}, even though we now realize that the
barrier between the ground state and that prolate shape isomeric state is
lowered by as much as 1 MeV by the inclusion of deformations that break the
axial symmetry, in particular in the heavier isotopes $^{204-208}$Pb. Rather
shallow oblate minima appear at an elongation of $q_2 \approx -0.45$ in the
$^{176-186}$Pb isotopes, with a small barrier of about 0.5 MeV that separates
them from the ground-state minimum. Since these minima turn out to be axially
symmetric they already appear clearly in Fig.\ \ref{Pt-Hg-Pb}. An additional
axially symmetric prolate minimum is predicted at $q_2\approx 0.35$ by our
calculations in $^{182-188}$Pb. It is located at about 1 MeV above the ground
state well, separated from it by a small barrier of about 0.5 MeV. Similarly,
small oblate minima are visible in Fig.\ \ref{Pt-Hg-Pb} for elongations in the
range $-0.3\!\lesssim\!q_2\!\lesssim\! -0.1$ in $^{184-194}$Pb, as was found
previously in some mainly self-consistent  approaches (see e.g.\
Refs.~\cite{BHR03,YBH13,NRR13,NVR02}). These turn out,  however, to disappear
when the non-axial degree of freedom is taken into  account, as this is clearly
seen in Fig.\ \ref{Pb-e21}. 

Let us refer the interested reader to a recent publication \cite{PNB19}
of our group, where shape isomers in nuclei of the three here studied isotopic 
chains have been extensively analysed and listed in a table in the 
appendix together with their energies with respect the ground-state, axial 
$Q_{20}$ and non-axial $Q_{22}$ quadrupole moments, and moments of inertia.


\section{Summary and conclusions}

In our study of isotopic chains of neutron deficient Pt, Hg and Pb nuclei with 
neutron numbers in the range $92 \leq N \leq 126$, a certain number of general
remarks can be made. 

First of all we have found that the left-right asymmetry degree of freedom does
not play any important role in any of the here considered nuclei, and this not
only around the ground state deformation, but throughout the here considered
range of deformations. However, this conclusion does not necessarily hold when
going to very large elongations (beyond $q_2 \!\approx\! 1$) as encountered in
the fission process.

Another important conclusion is that it appears absolutely essential to take
into account deformations that break the axial symmetry. It indeed appears in
all three of the here investigated isotopic chains that, at least for small
neutron numbers $92 \leq N \leq 106$, i.e.\ away from the $N \!=\! 126$ shell
closure, the energy landscape in the ($q_2,\,\eta$) plane is rather flat around
the ground-state deformation, which always stays rather close to the spherical
configuration, while a very pronounced spherical ground-state  minimum gradually
develops when approaching the magic number $N \!=\! 126$.  For the more neutron
deficient nuclei that present a very rich ($q_2,\,\eta$) landscape in the
vicinity of the spherical configuration, small differences  can decide whether
the ground-state deformation will turn out to be prolate or oblate, as this is
demonstrated when going from $^{186}$Pt, which is prolate deformed, to the
neighbouring $^{188}$Hg nucleus, by just adding a pair of protons, which then
turns out to have an oblate ground-state deformation.  A similar, even more
pronounced effect was already observed in in the Polonium  isotopes
\cite{NPB17}, on the other side of the $Z \!=\! 82$ shell closure,  when going
from $^{182}$Po with a prolate ground-state at $q_2 \!\approx\! 0.4$  with an
$\approx$ 1.5 MeV higher oblate isomer at $q_2 \!\approx\! -0.25$,  to
$^{192}$Po where the oblate minimum becomes the ground-state, before both  these
merge at $^{198}$Po with a ground-state which is spherically symmetric.

As far as the shape coexistence between prolate and oblate minima is concerned,
several candidates, in all three of the here studied isomeric chains, clearly
emerge: it is often away from the magic numbers, i.e.\ in regions where the
shell corrections are not that dominant, that such a phenomenon comes into play.
In the Platinum nuclei, we have found that the region $^{182-188}$Pt seems
particularly favourable, with minima at $q_2 \!\approx\! \pm 0.2$ separated by a
small barrier. In the Mercury isotopic chain it is in the $^{178-188}$Hg 
isotopes that a chance for shape-coexistence at small deformations should exist.
But the question of a possible shape-coexistence can, in fact, not be limited to
prolate versus oblate shapes, being more general,  as this has been  illustrated
through the ($q_2,\,\eta$) maps in the Lead isotopes, in particular in
$^{178-186}$Pb, where the landscapes in $\eta$ directions are very soft, thus
allowing for local minima, particularly visible in the landscapes of
$^{182-184}$Pb.

Future investigations of some additional isotopic chains in the $Z\approx82$ 
mass region, but also in other quite different mass regions are anticipated, in
order to probe, from a comparison with the experimental data, as that has  been
done here, the predictive power of our theoretical approach.
\vspace{-0.5cm}


\section*{Acknowledgements:}
\vspace{-0.2cm}
This work has been partly supported by the Polish-French COPIN-IN2P3
collaboration agreement under project numbers 08-131 and 15-149, as well as by
the Polish National Science Centre, grant No. 2016/21/B/ST2/01227.


\end{document}